# Astro2020 Science White Paper

# Extreme Plasma Astrophysics

**Thematic Areas:** ☐ Planetary Systems ☐ Star and Planet Formation
☒ Formation and Evolution of Compact Objects ☐ Cosmology and Fundamental Physics
☐ Stars and Stellar Evolution ☐ Resolved Stellar Populations and their Environments
☐ Galaxy Evolution ☒ Multi-Messenger Astronomy and Astrophysics

**Principal Author:**
Name: Dmitri A. Uzdensky
Institution: University of Colorado Boulder
Email: uzdensky@colorado.edu
Phone: 1-303-492-7988

**Co-authors:**
M. Begelman (JILA, Univ. Colorado Boulder), A. Beloborodov (Columbia Univ.), R. Blandford (KIPAC, Stanford Univ.), S. Boldyrev (Univ. Wisconsin Madison), B. Cerutti (Univ. Grenoble), F. Fiuza (SLAC, Stanford Univ.), D. Giannios (Purdue Univ.), T. Grismayer (Univ. Lisbon), M. Kunz (Princeton Univ.), N. Loureiro (MIT), M. Lyutikov (Purdue Univ.), M. Medvedev (Univ. Kansas), M. Petropoulou (Princeton Univ.), A. Philippov (Center for Computational Astrophysics, Flatiron Inst.), E. Quataert (Univ. California Berkeley), A. Schekochihin (Oxford Univ.), K. Schoeffler (Univ. Lisbon), L. Silva (Univ. Lisbon), L. Sironi (Columbia Univ.), A. Spitkovsky (Princeton,Univ.), G. Werner (Univ. Colorado Boulder), V. Zhdankin (Princeton Univ.), J. Zrake (Columbia Univ.), E. Zweibel (Univ. Wisconsin Madison).

# Extreme Plasma Astrophysics

D. Uzdensky et al.

(The words "extreme", "exotic", "traditional" are used here as technical terms with a specific meaning.)

## 1. Astrophysical Introduction

Relativistic compact astrophysical objects like neutron stars (NSs) and black holes (BHs) have fascinated both scientists and the general public for decades. Accreting BHs in galactic X-ray Binaries (XRBs) and active galactic nuclei (AGN) are powerful sources of high-energy (X-ray and gamma-ray) radiation. They also launch and accelerate ultra-relativistic collimated jets, which, in the case of AGN jets, propagate to enormous distances, carrying a substantial fraction of accretion energy; they dissipate some of this energy along the way and radiate it across the entire electromagnetic (EM) spectrum, from radio to very-high-energy TeV gamma rays. The rest of this energy is deposited into the medium surrounding the host galaxy, thus regulating the galactic-scale gas accretion and star formation processes. AGN jets are also believed to accelerate extremely energetic cosmic rays up to $10^{20}$ eV and may also be the source of extremely energetic (PeV) neutrinos, which makes them an attractive target for multi-messenger astronomical observations. Likewise, NSs provide us with a plethora of fascinating phenomena. Examples include: powerful pulses of radio, optical, X-ray, and gamma-ray radiation emitted with remarkable regularity by rapidly spinning magnetized pulsars; powerful X-ray emission from accreting NSs in NS XRBs; and extremely bright and short giant gamma-ray flares produced by magnetars---ultra-magnetized NSs with $\sim 10^{15}$ G magnetic fields. Newly born rapidly-rotating magnetars and BHs created in gravitational collapse of dying massive stars may both act as central engines of long Gamma-Ray Bursts (GRBs)---the most powerful explosions in the Universe since the Big Bang, while merging NSs are the most likely driver of short GRBs.

The last decade brought us even more amazing discoveries related to BHs and NSs: enigmatic very short and intense Fast Radio Bursts (FRBs); rapid gamma-ray flares in the Crab pulsar wind nebula (PWN) at GeV photon energies, apparently exceeding the standard synchrotron radiation-reaction limit; ultra-rapid (~10 min) TeV flares in blazar jets; giant BHs swallowing stars in Tidal Disruption Events (TDEs). And, just 1.5 years ago, a landmark observational breakthrough opened up the multi-messenger astronomy frontier---the detection of EM counterparts to the LIGO gravitational-wave (GW) event GW170817, caused by a merger of two NSs, across the EM spectrum, manifested as a short GRB, an optical kilonova, and a broadband afterglow. Another significant recent multi-messenger discovery is the IceCube detection of a very energetic neutrino coincident with a flaring blazar, providing the first direct evidence that AGN jets can accelerate hadrons to PeV energies. This never-ending flow of new observational discoveries and puzzling surprises keeps this field fresh and exciting. And we can reasonably expect that the rapid pace of such discoveries will continue, thanks to coming on line, in the near future, of new observational facilities such as ATHENA, LYNX, LISA, SKA, CTA, IXPE and others. It would be unprecedented if observations in the next decade do not present plasma astrophysics with completely new puzzles and challenges.

Most of what we know about these interesting phenomena comes from observations of EM emission at different wavelengths (sometimes augmented by multi-messenger neutrino and GW signals, as illustrated by the above examples), and all this light is produced by *plasma* *environments*. Thus, in order to understand the inner workings of these systems and how they convert gravitational and/or rotational energy into observable radiation, it is critical to have a solid, reliable understanding of how various basic plasma processes (such as various waves and instabilities, magnetic reconnection, turbulence, shocks) operate in these environments.

## 2. "Extreme" Plasma Astrophysics

To gain such an understanding, it is tempting to draw upon the huge body of theoretical knowledge that we now have about various plasma-physical processes in *traditional*---i.e., nonrelativistic, nonradiative, electron-ion---plasmas. This knowledge has been obtained over decades of intense research in space, solar, and laboratory plasma physics, via a combination of analytical theory, numerical simulations, laboratory experiments, and spacecraft measurements. It covers some of the most complex nonlinear multiscale phenomena, e.g., magnetized plasma turbulence and magnetic reconnection. While many puzzles in traditional plasma physics of course remain, and our understanding of it is not yet complete, the progress, especially in recent years, has been remarkable and impressive.

As a result, we now have a fairly good theoretical picture of how various plasma processes operate in *traditional* plasmas. However, the physical conditions in many plasma environments around NSs and BHs are *extreme* and quite different from those in traditional heliospheric and laboratory plasmas---not just quantitatively, but also qualitatively. They are richer, distinguished by several additional physical effects that would be considered "*exotic*" in traditional plasma physics. The most important such effects are:

- special-relativistic effects (relativistically hot plasmas and relativistic bulk motions);

- radiation-reaction effects (e.g., synchrotron or inverse-Compton radiative cooling);

- electron-positron pair creation;

- ultra-strong magnetic fields (QED effects such as 1-photon pair creation);

- general-relativistic (GR) effects.

Thus, one can say that while traditional plasma physics is fundamentally based on classical 19$^{th}$-Century physics (classic EM, statistical mechanics, fluid mechanics), the additional "exotic" new physics discussed here stems from the two main pillars of the 20$^{th}$-Century physics: theory of relativity and quantum mechanics (and hence quantum electrodynamics, QED).

Therefore, one cannot just blindly and straightforwardly apply traditional plasma-physics knowledge to the extreme astrophysical plasmas swirling around BHs and NSs; the above exotic effects can have a strong, but so far not well-understood, impact on even the classic fundamental plasma processes familiar to us from traditional plasmas. Furthermore, this new exotic physics can sometimes lead to completely new plasma processes. Thus, the work aimed at understanding extreme plasma-astrophysical phenomena offers an additional intellectual value of advancing plasma physics as a fundamental discipline.

## 3. Current Status of Extreme Plasma Astrophysics

Extreme plasma astrophysics is a vibrant and rapidly developing area of modern astrophysics and plasma physics. The intrinsic intellectual excitement of this discipline and its astrophysical applications drive modern theoretical, computational, and even experimental efforts.

The strong motivation to understand how the "exotic" relativistic, radiation, and QED physics affects various basic plasma processes in extreme plasma environments has by now been well recognized, especially in the high-energy astrophysics community. This recognition has led to a number of important, but so far isolated, efforts to study some of these effects. This includes both analytical studies and, especially in recent years, numerical simulations. Moreover, this is motivating a push for high-field QED-plasma laboratory experiments that take advantage of the impressive recent progress in high-intensity laser technology and test the limits of conventional accelerators.

On the *analytical front*, researchers are working to gauge the importance of various "exotic" physical processes in different regimes (usually, in concrete astrophysical contexts) and thus delineate the validity boundaries of traditional plasma physics [1]. They are also exploring how this "exotic" physics affects the familiar classic plasma processes (e.g., instabilities, shocks, reconnection, turbulence) [2-7]. Finally, they are casting their eyes towards new problems that arise only in extreme plasma astrophysics and do not appear in traditional plasmas, e.g., pair-production cascades [8].

Great progress is also being made on the *computational front*. Since the most interesting and relevant processes are complex and nonlinear already in traditional plasmas, and even more so in extreme plasmas, our ability to understand them greatly benefits from numerical simulation studies. Such studies have been picking up pace recently, revolutionizing our understanding of extreme plasmas. The rapid progress is in large part due to the development [driven both by astrophysical motivation and by laboratory high-energy-density-physics (HEDP) laser-plasma applications] of advanced numerical codes that go beyond the traditional kinetic, two-fluid, or magnetohydrodynamic (MHD) plasma descriptions and incorporate the exotic new effects.

Thus, while most traditional particle-in-cell (PIC) codes have long been fully relativistic, in the last few years several such codes (e.g., Zeltron, Tristan-MP, OSIRIS) have been augmented to include some of the more exotic effects, namely, synchrotron and inverse-Compton radiative cooling [9-12], pair creation [13-16], and even GR effects [17-20]. These developments are enabling, for the first time, truly ab-initio studies of important plasma-astrophysical phenomena such as the Crab Nebula gamma-ray flares [9,11], generation of pulsed high-energy emission in pulsar magnetospheres [10,14,18], coronae of accreting BHs, and pair-production cascades in pulsar [13-14,17] and black-hole magnetospheres [19-20], as well as generic, basic plasma processes such as magnetic reconnection [9,12,15]. Another important and promising avenue for numerical research in extreme plasma physics has been made possible in recent years by the advent and rapid development of radiative 3D GRMHD codes and their application to global simulations of accreting BHs and their jets [21-22]. In addition, researchers have developed ways to couple relativistic MHD codes to radiation and pair-production processes dynamically, leading to a significant progress in our understanding of radiative shocks in GRBs [23]. All these pioneering studies clearly demonstrate that rigorous numerical investigation of extreme plasma processes is now feasible and realistic, and this opens up broad avenues for future exploration.

The rapid progress in high-intensity laser technology is bringing the possibility to study some of these extreme plasma processes in the *laboratory experiments* for the first time and validate theoretical and numerical models. Relativistic $e^+e^-$ pair plasmas with densities up to $10^{16}$ cm$^{-3}$ have been recently produced by petawatt lasers exceeding $10^{21}$ W/cm$^2$ [24-26] and are starting to be used to probe collective effects associated with the interaction of these pair fireballs with ambient electron-ion plasmas [27-28]. The magnetic fields generated in these plasmas can exceed $10^9$ G and allow the study of magnetized environments relevant to accreting BHs, NS magnetospheres, and GRBs.

Although the intense laser fields are still orders of magnitude below the extreme fields found in rapidly-rotating NSs, the exotic strong-field QED processes that take place in NS magnetospheres could soon be accessible in the laboratory. This is because in the frame of an ultra-relativistic electron counter-propagating with a strong laser field, the field amplitude is Lorentz-boosted to values that can exceed the Schwinger critical limit ($4.4\times10^{13}$G). Above this threshold, the resulting high-energy photons can rapidly decay into $e^+e^-$ pairs, giving rise to a QED pair-cascade avalanche and to exotic plasma states dominated by the interplay of collective plasma processes and nonperturbative QED [29]. This is one of the most startling and exciting new plasma frontiers enabled by high-intensity lasers. Results from the first laser-based experiments approaching the Schwinger limit have just recently been reported [30-31]. In the next decade, experiments combining a petawatt laser with 10 GeV-class electron beams from linear accelerators (e.g., at SLAC) or multiple petawatt lasers (e.g., at ELI in Europe) may open a new window into the strong-field-QED regime of plasma processes relevant to extreme astrophysical environments.

## 4. Needs and Challenges

These great successes notwithstanding, extreme plasma astrophysics is still a very young, emerging discipline and has not yet reached full maturity or its full potential. The above-mentioned research efforts have so far been ad hoc, sporadic, and usually geared towards addressing a specific astrophysical problem, in a specific context. This differs starkly from traditional plasma physics where many basic scientific results have been accumulated and organized systematically as fundamental physics knowledge of general applicability, not necessarily tied to any specific context. In this respect, extreme plasma physics lags behind traditional plasma physics. Developing a general systematic knowledge base for extreme plasma physics is thus called for.

In particular, over the past decade or so, there has been great progress in our theoretical grasp of various fundamental processes in traditional collisionless plasmas---magnetic reconnection, shocks, turbulence, kinetic instabilities, particle acceleration, etc.,---which all had a strong impact on our understanding of systems like the Earth magnetosphere, solar wind, solar flares, the interstellar medium, the hot plasma in galaxy clusters, etc. However, the new theoretical concepts and ideas, as well as computational techniques, are yet to be transferred to the realm of extreme plasmas. There are certainly cases where the exotic physics (relativistic/radiation/QED) prevents one from a direct translation of these concepts from traditional to extreme plasma physics. At the same time, however, there are also other situations where such a generalization can be more or less straightforwardly carried out, it just has not been done yet. Doing this requires a strong dialog, a continuous exchange of knowledge and ideas between high-energy astrophysicists, who are well-versed in "exotic" relativistic and radiative physics, and experts in classical, traditional plasma physics.

## 5. Recommendations for Future Progress

To make the best use of these emerging opportunities and to build the field of extreme plasma astrophysics, we, as a community, need to combine and put to work our collective expertise in classic traditional plasma physics, high-energy astrophysics, numerical methods, radiative processes, relativistic fluid dynamics, and QED physics. Specifically, we recommend the following actions:

**(1) Training students** in both classic plasma physics and QED/radiative/relativistic physics.

**(2) Analytical theory**: establishing collaborations between traditional plasma theorists and high-energy astrophysicists to build theoretical models of basic plasma processes in extreme plasmas.

**(3)** Further development of new advanced **computational capabilities**, such as: fuller implementation of radiation, pair-creation/annihilation, and collisional (e.g., hadronic) processes, and GR effects in PIC codes; development of multi-scale, hybrid-physics codes (e.g., interface/integration between PIC and MHD modules within the same simulation); design of new numerical diagnostic and data analysis tools for digesting the results of simulation; development of standard benchmark test problems and protocols for code cross-comparison; and making the codes ready for computing at the upcoming exascale level.

**(4) Application** of these codes to classic fundamental plasma-physics problems in extreme plasma regimes: driven and decaying turbulence; magnetic reconnection; shocks; nonlinear development of instabilities [e.g., tearing, Weibel, magneto-rotational, kink, Kelvin-Helmholtz, Rayleigh-Taylor, mirror, firehose]; coherent radio emission; and pair-production cascades.

**(5)** Establishing and strengthening **connections to the HEDP** community and designing new experimental schemes for investigating extreme plasma physics in the lab.

## 6. Summary and Potential Impacts

In summary, *Extreme Plasma Astrophysics* --- a study of plasmas under extreme conditions, where special and general relativity, radiation, and QED processes such as pair creation and annihilation become important --- is an intellectually exciting emerging subfield of plasma physics with strong connections to astrophysics and HED physics. It is now experiencing a rapid progress enabled in part by powerful new numerical tools and experimental facilities, opening new research frontiers. This progress will have strong impacts on the broader scientific enterprise:

**(1)** Great benefit to **astrophysics**, leading to a much better, clearer understanding of many classes of extreme astrophysical objects and phenomena.

**(2)** Intellectual boost to **plasma physics**: understanding matter under extreme conditions, significantly expanding the realm of plasma physics to new physical regimes.

**(3) Training the next generation** of scientists who will have a combined expertise in astrophysics, traditional plasma physics, and ``exotic'' physics.